\documentclass[a4paper,11pt]{article}
\usepackage{pos}
\usepackage{graphicx}
\usepackage{subfigure}
\usepackage{caption}
\usepackage{subcaption}
\usepackage{lineno}
\graphicspath{ {images/} }

\title{Track-Like Event Analysis at the Baikal-GVD Neutrino Telescope}
\author[a]{V.M.~Aynutdinov}
\author[b]{V.A.~Allakhverdyan}
\author[a]{A.D.~Avrorin}
\author[a]{A.V.~Avrorin}
\author[c,d]{Z.~Barda\v{c}ov\'{a}}
\author[b]{I.A.~Belolaptikov}
\author[a]{E.A.~Bondarev}
\author[b]{I.V.~Borina}
\author[e]{N.M.~Budnev}
\author[l]{V.A.~Chadymov}
\author[f]{A.S.~Chepurnov}
\author[b,g]{V.Y.~Dik}
\author[a]{G.V.~Domogatsky}
\author[a]{A.A.~Doroshenko}
\author[c]{R.~Dvornick\'{y}}
\author[e]{A.N.~Dyachok}
\author[a]{Zh.-A.M.~Dzhilkibaev}
\author[c,d]{E.~Eckerov\'{a}}
\author[b]{T.V.~Elzhov}
\author[d]{L.~Fajt}
\author[l]{V.N. Fomin}
\author[e]{A.R.~Gafarov}
\author[a]{K.V.~Golubkov}
\author[b]{N.S.~Gorshkov}
\author[e]{T.I.~Gress}
\author[h]{K.G.~Kebkal}
\author[a]{I.V.~Kharuk}
\author[b]{E.V.~Khramov}
\author[b]{M.M.~Kolbin}
\author[i]{S.O.~Koligaev}
\author[b]{K.V.~Konischev}
\author[b]{A.V.~Korobchenko}
\author[a]{A.P.~Koshechkin}
\author[f]{V.A.~Kozhin}
\author[b]{M.V.~Kruglov}
\author[j]{V.F.~Kulepov}
\author[e]{Y.E.~Lemeshev}
\author[a,\dagger]{M.B.~Milenin}
\author[e]{R.R.~Mirgazov}
\author[b]{D.V.~Naumov}
\author[f]{A.S.~Nikolaev}
\author[a]{D.P.~Petukhov}
\author[b]{E.N.~Pliskovsky}
\author[k]{M.I.~Rozanov}
\author[e]{E.V.~Ryabov}
\author*[a]{G.B.~Safronov}
\author*[b,g]{D.~Seitova}
\author[b]{B.A.~Shaybonov}
\author[a]{M.D.~Shelepov}
\author[a]{S.D.~Shilkin}
\author[f]{E.V.~Shirokov}
\author[c,d]{F.~\v{S}imkovic}
\author[b]{A.E.~Sirenko}
\author[f]{A.V.~Skurikhin}
\author[b]{A.G.~Solovjev}
\author[b]{M.N.~Sorokovikov}
\author[d]{I.~\v{S}tekl}
\author[a]{A.P.~Stromakov}
\author[a]{O.V.~Suvorova}
\author[e]{V.A.~Tabolenko}
\author[b]{B.B.~Ulzutuev}
\author[b]{Y.V.~Yablokova}
\author[a]{D.N.~Zaborov}
\author[b]{S.I.~Zavyalov}
\author[b]{D.Y.~Zvezdov}

\affiliation[a]{Institute for Nuclear Research, Russian Academy of Sciences, Moscow, 117312, Russia}
\affiliation[b]{Joint Institute for Nuclear Research, Dubna, 141980, Russia}
\affiliation[c]{Comenius University, 81499 Bratislava, Slovakia}
\affiliation[d]{Czech Technical University in Prague, Institute of Experimental and Applied Physics, 11000 Prague, Czech Republic}
\affiliation[e]{Irkutsk State University, Irkutsk, 664003, Russia}
\affiliation[f]{Skobeltsyn Institute of Nuclear Physics, Moscow State University, Moscow, 119991, Russia}
\affiliation[g]{Institute of Nuclear Physics of the Ministry of Energy of the Republic of Kazakhstan, Almaty, 050032, Kazakhstan}
\affiliation[h]{LATENA, St. Petersburg, 199106, Russia}
\affiliation[i]{INFRAD, Dubna, 141981, Russia}
\affiliation[j]{Nizhny Novgorod State Technical University, Nizhny Novgorod, 603950, Russia}
\affiliation[k]{St.~Petersburg State Marine Technical University, St.~Petersburg, 190008, Russia}
\affiliation[l]{Moscow, free researcher}

\note[\textdagger]{Deceased.}
\emailAdd{grigorybs@gmail.com}
\emailAdd{diana.seitova.18@gmail.com}
\ConferenceLogo{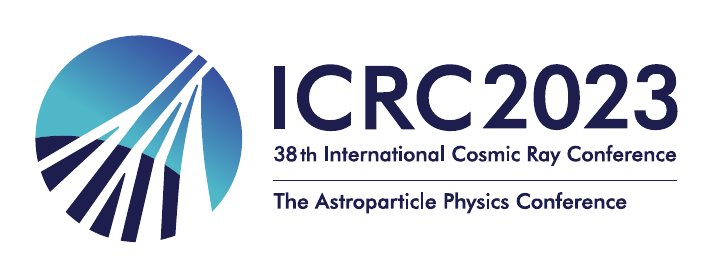}

\FullConference{%
38th International Cosmic Ray Conference (ICRC2023)\\
  26 July—3 August 2023\\
  Nagoya, Japan}

\begin{document}
\maketitle
{\bf Abstract. }Reconstructed tracks of muons produced in neutrino interactions provide the precise probe for the neutrino direction. Therefore, track-like events are a powerful tool to search for neutrino point sources. Recently, Baikal-GVD has demonstrated the first sample of low-energy neutrino candidate events  extracted from the data of the season 2019 in a so-called single-cluster analysis — treating each cluster as an independent detector. In this paper, the extension of the track-like event analysis to a wider data set is discussed and the first high-energy track-like events are demonstrated. The status of multi-cluster track reconstruction and that of the event analysis are also discussed.
\section{Introduction}
Baikal-GVD \cite{bib1} is the experiment in a construction phase located in the southern part of Lake Baikal. The depth of the lake in the installation area is approximately constant and is about 1366 m. The telescope has a modular structure and consists of independent detectors, so-called clusters. After the deployment campaign in winter 2023, the detector includes 12 standard clusters and one cluster with the experimental high-bandwidth DAQ. Each cluster has its own independent trigger and readout system and comprises 8 strings carrying optical modules (OMs) equipped with photomultipliers (PMTs) or control and readout electronics. The total number of deployed OMs with PMTs is 3456. A specialized module 30 m below the water surface controls the cluster readout system and is connected to the Shore Station by a hybrid electro-optical cable which provides the installation with electric power and enables data transmission. The data from the Shore Station are transferred to JINR (Dubna) where they are processed and stored. The stored data are divided into a single-cluster event set (individual events for each cluster) and a multi-cluster event set (combined events in which the coincidence of triggers on different clusters was found — in a time window defined by the maximum muon propagation time between these clusters). 

%The main goal of the experiment is to study the astrophysical neutrino flux discovered by the IceCube telescope in 2013 \cite{bib3}.
The main goal of the experiment is to study the astrophysical neutrino flux. The neutrino interaction with matter in the vicinity of the detector is accompanied by a cascade of charged particles and/or a muon that propagates over large distances keeping the direction of the neutrino with a good accuracy (which improves with increasing energy). The direction and energy of the muon can be estimated using the times and integrated charges of the pulses registered in the PMTs along the track. The angular resolution for track events is at the level of 0.5° for muons with a sufficiently long track in the sensitive detector volume, which is substantially better than the typical resolution for cascades. At the same time, the energy resolution is much worse than that for cascade events \cite{cascade}. A good directional resolution makes the track analysis especially important for the search for neutrino point sources. 
%\begin{figure}[h]
%\includegraphics[width=0.5\textwidth]{images/gvd.png}
%\caption{\bf {angular resolution graph.}}
%\end{figure}
%\begin{figure}[h]
%\includegraphics[width=0.5\textwidth]{images/gvd.png}
%\caption{\bf {energy resolution graph.}}
%\end{figure}
%\paragraph{}

The majority of neutrinos reconstructed by the detector are produced in the Earth’s atmosphere. The spectrum of atmospheric neutrinos decays rapidly as energy increases. At particle energies of several tens of TeV, the flux of astrophysical neutrinos starts to dominate. Both single-cluster and multi-cluster events generated by the data processing system are important for the track analysis. Single-cluster events ensure the sensitivity of the telescope to ascending neutrinos with zenith angles greater than ~120°. Multi-cluster events provide the sensitivity to nearly horizontal events which are important at high energies (> O(500 TeV)) since the Earth loses transparency to neutrinos in this energy range. The analysis of track-like events is aimed at an efficient and precise neutrino reconstruction in the ~100 GeV–multi-PeV energy range. In this paper, the recent developments in the track-like event analysis are discussed. 

\begin{figure}[hbtp!]
%\begin{center}
\centering
%   \begin{subfigure}[b]{0.43\textwidth}
%	   \centering
	   \includegraphics[width=6.5 cm]{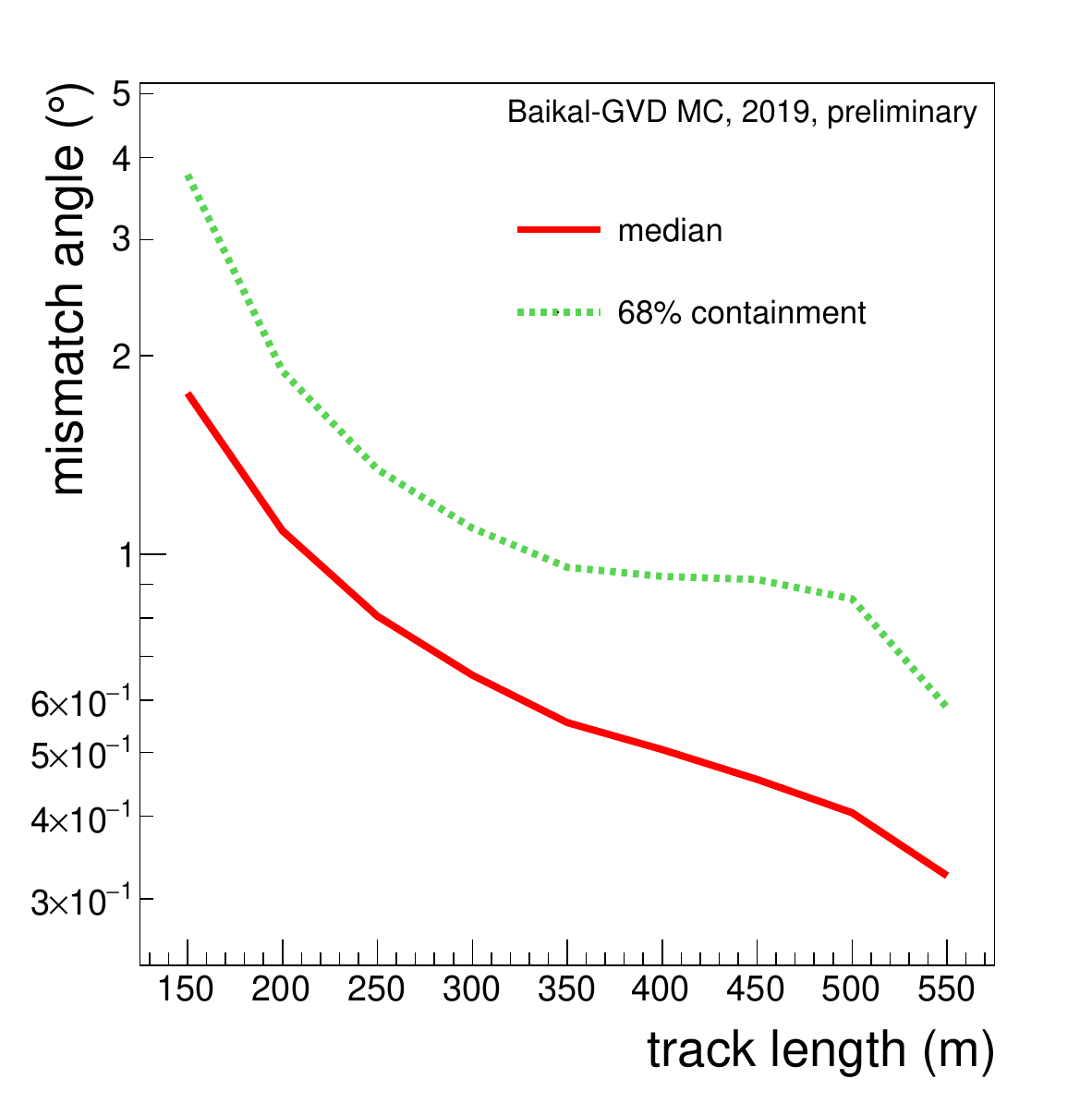}
%	   \caption{a} 
%    \end{subfigure}
%\hspace{0.5 cm}
%    \begin{subfigure}[b]{0.43\textwidth}
%	   \centering
%	   \includegraphics[width=6.1 cm]{images/angular_resolution.pdf}
%	   \caption{b}
%    \end{subfigure}
\caption{Angular resolution for single-cluster reconstruction of muons with the energy between 10 TeV and 1000 TeV  in the centre of the detector\label{fig1}}
%\end{center}
\end{figure}

\section{Muon track reconstruction algorithm and its performance}
Muon tracks are reconstructed in two stages. At the first stage, PMT pulses (hits) generated by Cherenkov light emitted by muon are selected while noise pulses from lake water luminescence, PMT dark current or afterpulses are suppressed. At the second stage, the direction, energy, and various track quality parameters (e.g. fit convergence, value of minimisation function) are reconstructed. 
Noise suppression for single-cluster reconstruction is performed with the efficient algorithm based on the directional causality criterion and fast causally connected pulse search algorithm \cite{bib9}. The celestial sphere is scanned with an adjustable step (5–10°). For each direction, a clique of pulses which fall within a direction-dependent time window is identified. The direction with a sufficiently large number of selected pulses and the minimum value of the quality function is chosen as a preliminary one, and the corresponding set of pulses is used to accurately reconstruct the track parameters. This approach retains a large fraction of signal pulses in a large energy range keeping the noise pulse contamination at the level of few \%. The multi-cluster reconstruction has a dedicated noise suppression algorithm. The track is approximated using two pairs of pulses identified by the trigger system \cite{bib1} in different clusters, and other pulses are collected using time window with respect to this approximation. The track fit procedure is similar to the single-cluster analysis. The track fit is performed using pulse collection and the preliminary track direction as found by the noise suppression algorithm. The track direction, position, and time are found by minimising the quality function consisting of a $\chi^2(t)$ term and a term proportional to the sum of products of pulse charge and the distance from the track. Tens of track quality parameters and the muon energy are reconstructed at this stage. The energy is reconstructed using mean muon energy loss along the track estimation as a proxy. The track direction measurement precision depends on the track length and varies from $\sim1.5^{\circ}$ for short tracks to below 0.5$^{\circ}$ for long nearly vertical tracks (Fig.\ref{fig1}). The energy resolution depends on energy and varies between the factors of 3 and 3.5.
 The directional resolution for the multi-cluster analysis is better due to larger track lengths and approaches 0.2$^{\circ}$ for the longest tracks.

%\begin{equation}
%Q=\sum{(t-t_{th})^2 \over \sigma^2} + {0.3({N_{hits}-6})\over Q} \sum{a_{0}q \sqrt{d_{1}^2+d^2} \over \sqrt{a_{0}^2 + q^2}}
%\end{equation}

\section{Progress in neutrino candidate selection}
Most of the events passing through the muon reconstruction are atmospheric muon bundle events. Events from atmospheric muons, incorrectly reconstructed as ascending ones, form the background to atmospheric neutrinos exceeding the signal by a factor of $10^2-10^4$ depending on the zenith angle. This background is suppressed by applying cuts on various track parameters or by using the boosted decision tree (BDT) technique. 

The first set of candidate events in the single-cluster analysis was selected using the so-called fast single-cluster event reconstruction algorithm which was applied to the data from the first three months of the year 2019. As a result, a set of 44 neutrino candidate events was selected in a cut-based analysis \cite{bib6}. The obtained event set is dominated by atmospheric neutrinos with an average energy of ~500~GeV.  

The improved method for the hit finding  \cite{bib9} combined with the BDT-based suppression of the misreconstructed background allowed enhancing the neutrino yield from the same dataset in the single-cluster analysis by a factor of 2 \cite{bib10}. 
%The distribution of tracks along the zenith angle obtained using the described algorithm is shown in Fig. 2(a).
 %The BDT was trained on sets of MC signal and background events with a reconstructed zenith angle > 120°. It was found that the threshold on the BDT classifier value (response) of 0.25 (see Fig. 2(b)) completely suppresses the background for the telescope live time corresponding to the first three months of the 2019 season, while retaining 70\% of the signal. As a result of applying this classifier to the data, a set of 106 candidate events was identified in 326 days equivalent single-cluster livetime. The expected number of MC candidate events for atmospheric neutrinos is 81.2. The detected discrepancy of ~30\% is being studied. As a result of the work done, the sensitivity of the telescope to low-energy neutrinos was improved by about a factor of two compared to \cite{bib6}. 

\begin{figure}[hbtp!]
\begin{center}
%\centering
%\begin{subfigure}{0.43\textwidth}
%	\centering
	\includegraphics[width=2.5 cm]{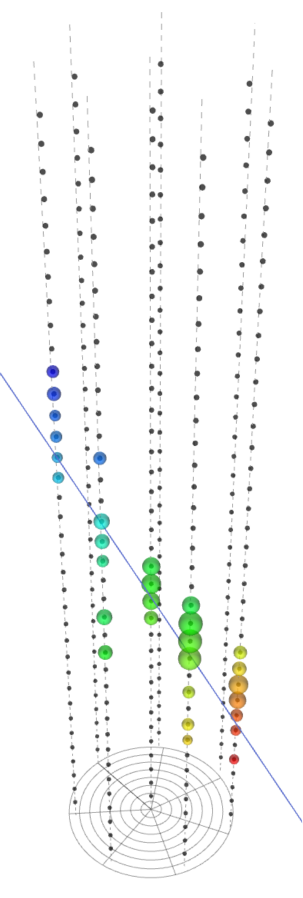}
%	\caption{ } 
%\end{subfigure}
    \hspace{2.5 cm}
%\begin{subfigure}{0.43\textwidth}
%	\centering
	\includegraphics[width=6.5 cm]{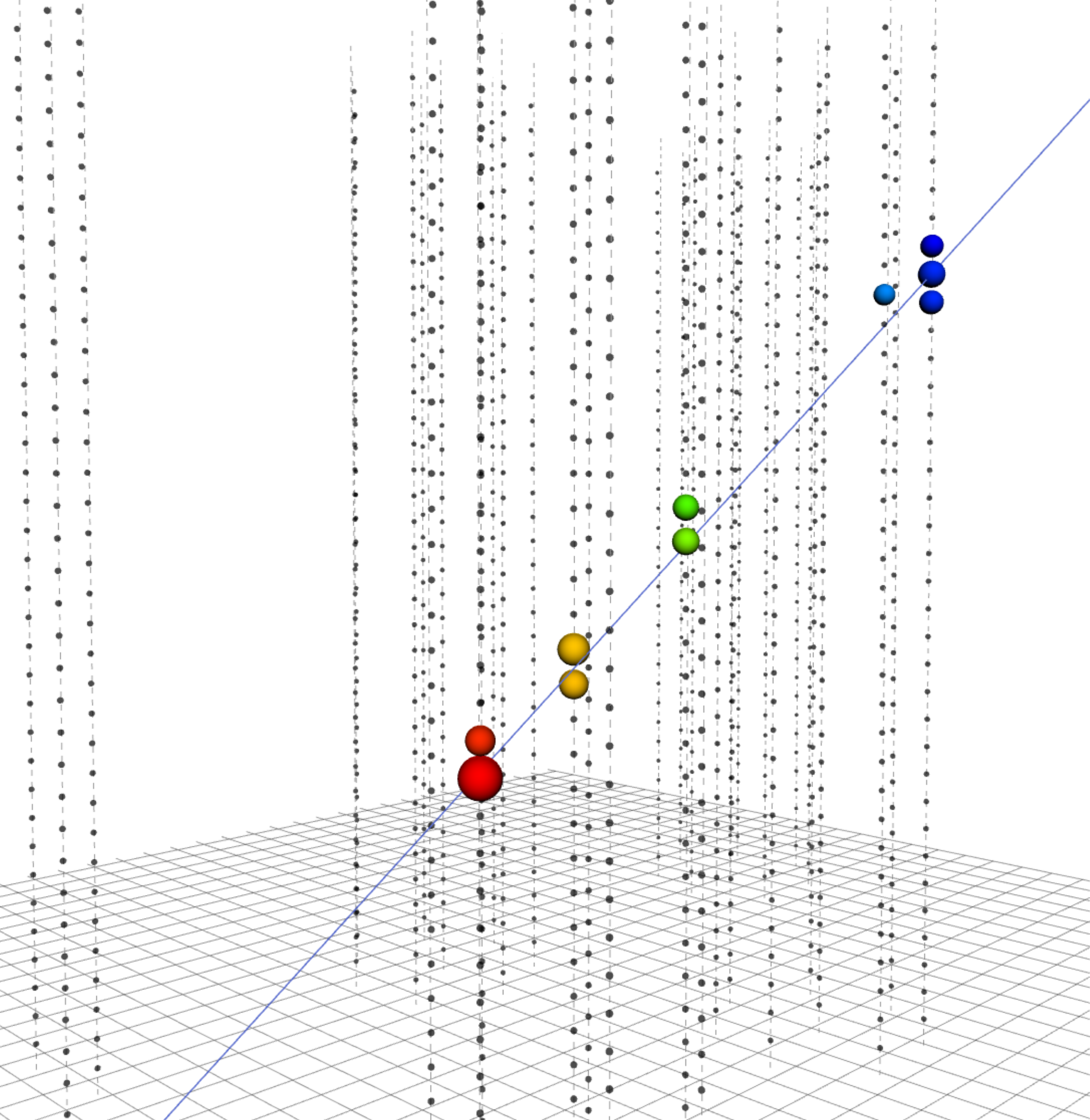}
%	\caption{ }
%\end{subfigure}
\vspace{0.5 cm}
\caption{ {\bf Left:} The 100-TeV upgoing muon neutrino candidate selected in the single-cluster analysis with a high probability of astrophysical origin. The estimated signalness of the event is~>90\%. The event was registered in autumn 2019. {\bf Right:} The multi-cluster neutrino event candidate registered in summer 2019 \label{fig2}}
\end{center}
\end{figure}

Single-cluster reprocessing of the full 2019 dataset with the improved reconstruction and neutrino selection methods allowed to further increase the low-energy neutrino candidate dataset and also identify the first high-energy track-like neutrino candidate events. Since the analysis is still in progress, we present preliminary characteristics of one of such events (Fig.\ref{fig2}, left). An upgoing muon with the median estimate of the energy corresponding to 103.4 TeV was detected in autumn of 2019. The number of hits selected by the hit selection algorithm is 30, the reconstructed zenith angle, $\theta$, of the event is 153.4$^{\circ}$, and the visible track length is 332.4 m. The angular resolution corresponding to a 68\% containment interval is 0.67$^{\circ}$. The probability of the event to be of astrophysical origin (signalness) was estimated taking into account contributions from atmospheric muon bundles, atmospheric neutrinos, including prompt neutrino contribution, and using the astrophysical neutrino spectrum with $\gamma_{astro}\sim$-2.36 according to the recent IceCube extraction of the spectrum for track-like events \cite{bib11}. The signalness of the event was estimated at >90\%.

Multi-cluster track-like events allow to fully unlock the potential of the telescope's angular accuracy. Presently, the multi-cluster analysis is under development. The preliminary cut-based analysis optimised on MC and applied to a fraction of the 2019 data resulted in the detection of the first multi-cluster track-like neutrino candidate (Fig.~\ref{fig2}, right). The selected event is a two-cluster event with 10 hits distributed among 5 strings and a track length of ~399 m. The reconstructed zenith angle of the track is 125.6$^{\circ}$. The event was recorded on July 6, 2019.  

\section{Conclusions}
The Baikal-GVD track-like event analysis is progressing towards a higher sensitivity and precision. The muon reconstruction techniques developed so far make it possible to achieve an angular resolution of well below 0.5° for sufficiently long tracks. The single-cluster analysis of the full 2019 dataset resulted in the detection of the first high-energy muon events, the most spectacular one being the upgoing muon event with an energy of 102 TeV and signalness of > 90\%. The multi-cluster track-like event analysis is being developed, and the first multi-cluster neutrino candidate has been registered.

%% Full authors list (ONLY FOR COLLABORATIONS)
%\clearpage
%\section*{Full Authors List: Baikal-GVD Collaboration}
%
%\noindent \textbf{Note comment afterwards:} Collaborations have the possibility to provide an authors list in xml format which will be used while generating the DOI entries making the full authors list searchable in databases like Inspire HEP. \\
%
%\scriptsize
%\noindent
%V.A.~Allakhverdyan, A.D.~Avrorin, A.V.~Avrorin, V.M.~Aynutdinov, Z.~Barda\v{c}ov\'{a}, I.A.~Belolaptikov, E.A.~Bondarev, I.V.~Borina, N.M.~Budnev, V.A.~Chadymov, A.S.~Chepurnov, V.Y.~Dik, G.V.~Domogatsky, A.A.~Doroshenko, R.~Dvornick\'{y}, A.N.~Dyachok, Zh.-A.M.~Dzhilkibaev, E.~Eckerov\'{a}, T.V.~Elzhov, L.~Fajt, V.N. Fomin, A.R.~Gafarov, K.V.~Golubkov, N.S.~Gorshkov, T.I.~Gress, K.G.~Kebkal, I.V.~Kharuk, E.V.~Khramov, M.M.~Kolbin, S.O.~Koligaev, K.V.~Konischev, A.V.~Korobchenko, A.P.~Koshechkin, V.A.~Kozhin, M.V.~Kruglov, V.F.~Kulepov, Y.E.~Lemeshev, M.B.~Milenin, R.R.~Mirgazov, D.V.~Naumov, A.S.~Nikolaev, D.P.~Petukhov, E.N.~Pliskovsky, M.I.~Rozanov, E.V.~Ryabov, G.B.~Safronov, B.A.~Shaybonov, M.D.~Shelepov, S.D.~Shilkin, E.V.~Shirokov, F.~\v{S}imkovic, A.E.~Sirenko, A.V.~Skurikhin, A.G.~Solovjev, M.N.~Sorokovikov, I.~\v{S}tekl, A.P.~Stromakov, O.V.~Suvorova, V.A.~Tabolenko, B.B.~Ulzutuev, Y.V.~Yablokova, D.N.~Zaborov, S.I.~Zavyalov, D.Y.~Zvezdov
%first.author$^1$, 
%second.author$^2$, 
%third.author$^3$ % .... more names
%and 
%last.author$^{n}$ \\
%
%\noindent
%$^1$first.affiliation.
%$^2$second.affiliation. % .... more affiliation
%$^{m}$last.affiliation.

\end{document}